\documentclass[11pt,twoside]{article}

\usepackage{asp2006}
\usepackage{epsf}

\begin{document}
\title{Emission-Line versus Continuum Correlations in Active Galactic Nuclei}   %%% Fill in title
\author{Joseph C. Shields}   %%% Fill in author names
\affil{Physics \& Astronomy Department, Ohio University, Athens, OH 45701 USA}    %%% Fill in author affiliations

\begin{abstract} %%% Abstract to run on from here.
The Baldwin Effect, a negative correlation between emission-line equivalent
width and luminosity in active galactic nuclei, is still of interest as a 
diagnostic of accretion physics nearly thirty years after its discovery.  
This review examines recent developments in the study of correlations 
between line and continuum emission in AGNs, as measured both in ensembles
and in individual sources.
\end{abstract}

%%% MAIN BODY OF TEXT GOES HERE. CONSULT "INSTRUCTIONS FOR AUTHORS USING
%%% LATEX2E MARKUP", SECTIONS 2.3-2.6 FOR HELP WITH EQUATIONS, FIGURES,
%%% AND TABLES.
\vspace{-0.1in}
\section{Introduction}   %%% Top level section head (remove "%" symbol)

A simple but important observational question in the study of active
galactic nuclei (AGNs) is whether emission-line luminosities scale in
proportion to continuum luminosity for the central source.  This topic
has been investigated in multiple studies over the past three decades.
An early and influential paper on this subject was published by
\citet{Baldwin77}, who reported a strong, negative correlation between
the restframe equivalent widths for the ultraviolet (UV) lines
(particularly \ion{C}{iv} $\lambda$1549) and continuum luminosity in
quasars.  \citet{Carswell78} referred to this trend as the {\em
  Baldwin Effect}, a label thereafter adopted by the AGN community.
The Baldwin Effect can be expressed in terms of line luminosity or
equivalent width (EW), and both descriptions commonly appear in the
literature.  An illustration of what the Baldwin Effect means for
quasar spectra is shown in Figure 1, which shows a set of composite 
spectra from \citet{Dietrich02} for sources binned by luminosity.

\begin{figure}[t]
\plotfiddle{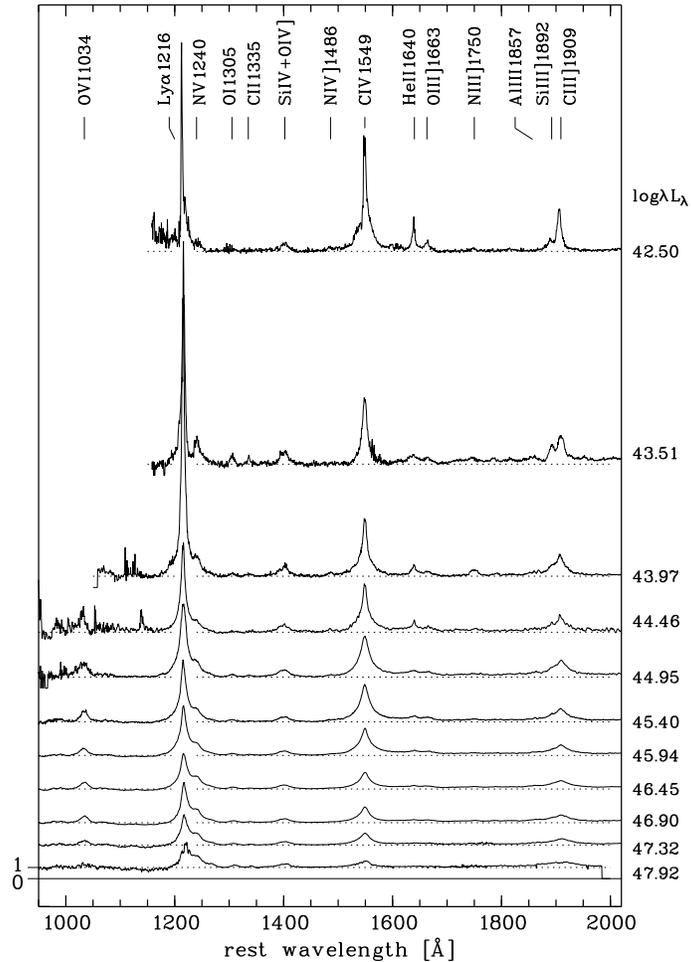}{4.8in}{0}{50}{50}{-145}{-20}
\caption{Normalized composite spectra for QSO subsamples binned by luminosity
[$\Delta\log\lambda L_\lambda(1450 {\rm \AA} ) = 0.5$ dex].  Individual spectra
are shown on a common vertical scale, shifted for display purposes
\citep{Dietrich02}.}
%\vspace{-0.2in}
\end{figure}

Nearly thirty years after Baldwin's original paper, the correlation he
reported still remains interesting in relation to BLR physics.
Opportunities to study line versus continuum correlations have
improved dramatically in recent years, thanks to better and more
abundant data.  Something I want to emphasize in this review, however,
is that the context for understanding the Baldwin Effect has also
grown, thanks to the emergence of sophisticated tools such as
Principal Components Analysis, and techniques for estimation of black
hole masses.  While emission-line EWs may seem rather primitive
diagnostics, they retain the virtue of being easily measurable,
including in cases where flux calibration is sometimes difficult, such
as large optical surveys with fiber-fed spectrographs.

Some clarification of the scope of this review is warranted given the
varied phenomena that have been tagged with the Baldwin Effect label.
The original trend is a correlation between emission lines and
continuum luminosity measured from single-epoch observations of
multiple QSOs.  Studies of individual, variable AGNS have also
revealed negative correlations between emission-line EWs
and continuum luminosity ($L$), which \citet{Pogge92} dubbed the
``intrinsic'' Baldwin Effect to distinguish it from the original
ensemble (or ``global'') effect.  While most of the early studies of
the Baldwin Effect focused on UV/optical broad emission lines, more
recent work has also investigated in some detail correlations between
narrow emission-line and continuum luminosities.  Improvements in
X-ray technology have permitted detailed studies of the Fe K$\alpha$
line, and there have been a number of claims that this feature shows
an ``X-ray Baldwin Effect''.  But claims of a Baldwin Effect have
moved beyond AGNs with reports of such a trend in the optical emission
lines for Wolf-Rayet stars \citep{Morris93} and dwarf novae
\citep{Long05}\footnote{To confuse things further, searching in the
  NASA Astrophysics Data System on the words ``Baldwin Effect'' will
  also turn up references to the computer science literature,
  referring to an idea in evolutionary biology originated by
  \citet{Baldwin96}.}.  A key point to keep in mind is that the
Baldwin Effect is simply an observational correlation; the physics
underlying the correlation may be very different in each of these
cases.  In this review I will be restricting the discussion to AGNs,
and even within these sources the different Baldwin trends (broad
lines, narrow lines, intrinsic, X-ray) may have disparate origins.  I
will also emphasize recent developments; a review of the older
literature on the subject and detailed discussion can be found in
\citet[hereafter OS]{Osmer99}.

\section{The Ensemble Baldwin Effect}

\subsection{What Do We Know?}

By now several aspects of the ensemble Baldwin Effect are well
determined.  A first point that has generated some controversy in the
past is simply that {\it it exists}.  Studies following Baldwin's
initial report generally found greater scatter and sometimes little
evidence of any trend.  It now seems that significant scatter is
typical of most AGN samples, and the key to finding the correlation is
having a sufficiently large span of luminosity in the sample; an
excellent illustration of this point can be found in the study by
\citet{Kinney90}, which was among the first to probe AGNs covering 6
orders of magnitude in UV continuum luminosity.

A second aspect of the Baldwin Effect that now seems firmly
established is the existence of {\it different slopes for different
  lines}.  The variation is not random (Figure 2); lines originating
from higher ionization species display steeper slopes in the EW versus
$L$ diagram (e.g., \citeauthor{Espey99}\citeyear{Espey99};
\citeauthor{Dietrich02} \citeyear{Dietrich02}).

\begin{figure}[t]
%\vspace{0.2in}
\plotfiddle{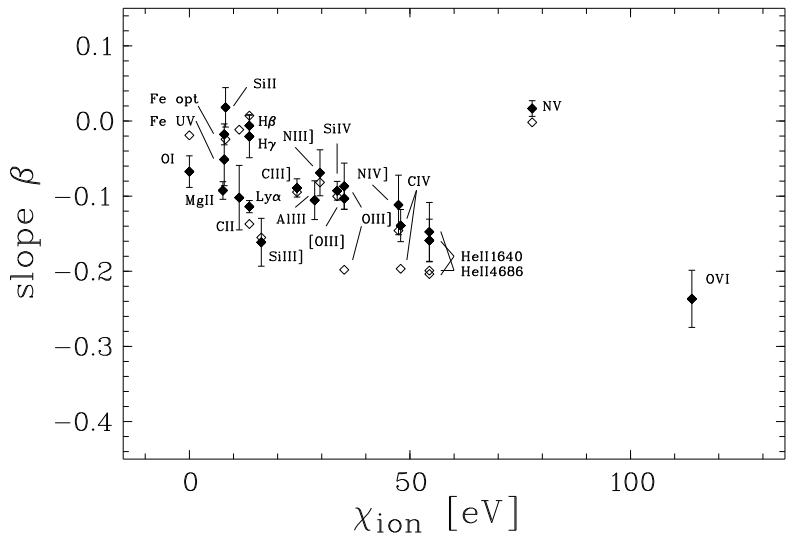}{2.5in}{270}{40}{40}{-160}{220}
\caption{Slope $\beta$ of the best linear fit in the log EW vs.
log $L$ plane for emission lines of the indicated species, plotted as a
function of the ionization energy $\chi_{ion}$ required to create the
ion \citep{Dietrich02}.  Open symbols represent slopes obtained using
only high $L$ sources (see \citeauthor{Dietrich02} for details).}
%\vspace{-0.2in}
\end{figure}

A few lines depart from these trends and merit special mention.
Multiple studies have shown that the \ion{N}{v} $\lambda$1240 EW
shows little if any dependence on $L$ -- in other words, no
Baldwin Effect, which is particularly noteworthy given the large
ionization potential for this ion \citep[and references therein]
{Dietrich02}.  The most likely explanation for its exceptional behavior
is a nucleosynthetic one, since secondary nitrogen enrichment will cause
the N/H ratio to scale as the square of the global metallicity; if
higher luminosity sources are also characterized by higher metallicity,
the selective enhancement of nitrogen in luminous sources may compensate
for the processes that would otherwise diminish \ion{N}{v} emission
as expected from the Baldwin relations \citep{Hamann93,Korista98}.

An additional line of special interest is H$\beta$ $\lambda$4861.  Most
studies have shown little luminosity dependence in the Balmer lines,
but recent work with large samples has demonstrated a weak {\it
inverse} Baldwin Effect for H$\beta$: $EW \propto L^{0.2}$ based on
the 2dF+6dF surveys \citep{Croom02}, and $EW \propto L^{0.1}$ using the
Sloan Digital Sky Survey (\citeauthor{Greene05}\citeyear{Greene05}; 
see also \citeauthor{Netzer07}\citeyear{Netzer07}).  The tight relationship
between line and continuum luminosity means that H$\beta$ can be used
as a surrogate for the optical or total luminosity.  Since the
velocity width of the same line serves as a kinematic tracer of
the BLR, this single feature provides the necessary ingredients to
estimate the black hole mass $M_\bullet$ in a given source 
(\citeauthor{Greene05}\citeyear{Greene05}; see also X.-B. Wu, this 
conference).

\subsection{Explanations}

Several physical explanations have been proposed to account for the
Baldwin Effect, and its dependence on ionization.  A hypothesis that
has gained considerable support is that the continuum shape is
luminosity-dependent such that more luminous objects have softer
UV/X-ray spectra, resulting in reduced ionization and photoelectric
heating in the BLR gas \citep[e.g.,][]{Netzer92,Korista98}.  In that
case the equivalent widths of emission lines diminish at higher
luminosity with the strongest effect expected for high-ionization
lines, as observed.  Independent analyses of AGN spectral energy
distributions confirm a luminosity dependence in the continuum that it
is at least qualitatively consistent with this picture
\citep[e.g.,][]{Strateva05}.

A question that merits careful consideration is whether the Baldwin
Effect is actually driven by a relationship between EWs and another
fundamental parameter that happens to be correlated with $L$ in
typical AGN samples.  Specific examples of such a parameter would
include redshift $z$, the Eddington ratio $L/L_{Edd}$, and
$M_\bullet$.  Distinguishing which of these properties is a key driver
in the correlations is not easy.  Testing the dependence of EW on $L$
and $z$ independently requires a sample that fills out the $L - z$
plane; the hard part, of course, is obtaining measurements of low $L$
sources at high $z$.  Progress in this matter has been demonstrated
recently by \citet{Dietrich02}, who assembled a large sample from various
existing surveys that allows measurement of EW as a function of $L$ at
constant $z$, and EW as a function of $z$ at constant $L$.  The results
convincingly demonstrate that luminosity dominates the trend, and any
dependence on $z$ is weak in comparison.

Several recent studies have investigated the question of whether
$L/L_{Edd}$ is actually the driving parameter for the Baldwin Effect.
Estimates of $L/L_{Edd}$ require values of $M_\bullet$, which in turn
are obtained from luminosities and emission-line velocity widths,
using calibrations from reverberation mapping. \citet{Baskin04} found
that EW(\ion{C}{iv}) for the PG quasars shows a tighter relationship
with $L/L_{Edd}$ than with $L$.  Using a different sample,
\citet{Netzer04} likewise found evidence for a stronger correlation
with the Eddington ratio, but using EW(H$\beta$), where the very
existence of a conventional Baldwin Effect appears inconsistent with
other studies (see above), a result \citeauthor{Netzer04} ascribe to
possible sample incompleteness.  Via more circuitous arguments,
\citet{Bachev04} also claim evidence for $L/L_{Edd}$ as the more
fundamental parameter underlying the behavior of \ion{C}{iv} in
particular.  \citet{Warner04} have investigated other lines in the
\citet{Dietrich02} sample, and find evidence for significant
correlations in many cases between EW and $L/L_{Edd}$, but with slopes
that are rather different from those seen in the conventional Baldwin
diagrams (e.g., a significant Baldwin Effect for \ion{N}{v} and an
inverse trend for \ion{O}{vi} $\lambda$1034).

Assessing the relative importance of $L$, $L/L_{Edd}$, and $M_\bullet$
in driving the Baldwin correlations is complicated by the fact that
these quantities can be highly correlated in typical samples.  To
address this problem, \citet{Warner06} have most recently used
composite spectra built from subsamples to separate the dependencies
on these variables.  Their results illustrate spectral variations that
result as $L/L_{Edd}$ is incremented while $M_\bullet$ remains
constant, and as $M_\bullet$ is incremented while $L/L_{Edd}$ remains
constant.  In the former case, little change in equivalent widths is
seen as $L/L_{Edd}$ varies, while in the latter case, a strong
dependence on $M_\bullet$ appears.  Their sample overall shows the
classical Baldwin dependence on $L$, but the spectral changes diminish
substantially when $L$ is varied for a subsample restricted to a
narrow range of $M_\bullet$.  In contrast, dramatic line variations
are seen as $M_\bullet$ is varied while $L$ remains fixed.  In
summary, \citeauthor{Warner06} provide strong evidence that
$M_\bullet$ is likely to be the fundamental parameter driving EW
variations that appear as the Baldwin Effect.  This finding has an
attractive physical consistency in that more massive black holes are
expected to have cooler accretion disks that in turn will produce
softer continua, in qualitative accord with the ionization dependence
of the Baldwin Effect (e.g., \citeauthor{Netzer92}\citeyear{Netzer92};
\citeauthor{Wandel99} \citeyear{Wandel99}).

\subsection{New Approaches}

By now, a variety of other trends and correlations have emerged out of
the expanding spectral data for AGNs, and to genuinely understand the
Baldwin Effect and its implications, we must integrate it into this
larger context.  Fortunately we also now have tools that allow us in
principle to achieve this goal, where in particular I am referring to
Eigenvector or Principle Components Analysis (PCA; e.g.,
\citeauthor{Boroson92} \citeyear{Boroson92};
\citeauthor{Francis92}\citeyear{Francis92};
\citeauthor{Shang03}\citeyear{Shang03}).  In a multi-dimensional space
of measured spectral parameters, PCA identifies eigenvectors that
trace the maximal variance in the system, thereby reducing its
dimensionality if some of the parameters are correlated.  Expressed
another way, PCA makes it possible to find patterns in the spectra of
large AGN samples, describing in an objective way which among many
spectral properties correlate, and how.  The hope is that once such
patterns are identified, they can point us to understanding what
fundamental physical properties are ultimately responsible for
defining the observed properties of AGNs.

One of the more comprehensive PCA analyses for AGNs that encompasses
both the restframe UV and optical bandpasses is that of
\citet{Shang03}.  Their study employs ``spectral PCA'' such that the
measured parameters are values of flux as a function of wavelength in
normalized spectra; the resulting eigenvectors then have the
appearance of spectra.  These eigenvectors are labeled as Spectral
Principal Component (SPC) 1, 2, 3, etc. according to their dominance
in accounting for the variance in the input
dataset. \citeauthor{Shang03} argue that the Baldwin Effect is closely
linked to their SPC1, which shows prominent line cores and diminishes
in strength as source luminosity increases.  A separate SPC, their
SPC3, is linked to the ``Eigenvector 1'' (EV1) identified in optical spectra
by \citet{Boroson92}, which those authors argue is driven by either
$L/L_{Edd}$ or $M_\bullet$.  The upshot is that PCA studies do not
draw a strong connection between the Baldwin Effect and either 
Eddington ratio or $M_\bullet$ as key underlying parameters.

An informative way to probe this issue further is to examine the
behavior of Narrow-Line Seyfert 1 (NLS1) galaxies in the Baldwin
diagrams.  NLS1 galaxies display broad H$\beta$ with relatively small
velocity widths, and weak forbidden lines as well as distinctive X-ray
properties.  Their characteristics overall place them at one extreme
of the EV1 sequence and there is considerable belief that they are in
a high accretion state with large $L/L_{Edd}$.  \citet{Leighly04} have
examined the behavior of NLS1s in the Baldwin diagrams for UV lines
and find the NLS1s are systematically offset to lower EWs at a given
continuum luminosity, relative to broad-line AGNs considered more generally.
We might expect the EW trends to show more homogeneous behavior when
EW is plotted against $L/L_{Edd}$, since the expectation is that the
NLS1s will slide to the right-hand side of the diagram where EWs are low.
But when this exercise is actually carried out, as shown by \citet{Warner04},
the outcome is not so clean.  For some emission features NLS1s align 
with other AGNs in the EW vs. $L/L_{Edd}$ diagram, but for others the
NLS1s fall {\it sometimes below and sometimes above} the full ensemble 
trend -- and notably, the NLS1s do not show the largest $L/L_{Edd}$ values
among their sample.  Does the latter statement imply that \citet{Warner04}
systematically mis-estimated $L/L_{Edd}$ for NLS1s, due perhaps to their
use of \ion{C}{iv} rather than the better-calibrated H$\beta$ line in
deriving $M_\bullet$?  Even if we suppose this
to be the case and slide the NLS1 results horizontally in the EW vs.
$L/L_{Edd}$ diagrams, these objects remain outliers from the main trend,
often with disagreement in different directions for different lines.  As
noted by \citet{Warner04}, the results suggest that one or more additional
parameters beyond $L/L_{Edd}$ are important in defining the spectral
characteristics of NLS1s.

To summarize, conventional correlation analyses and PCA to date provide 
mixed indications of a major role for $L/L_{Edd}$ or $M_\bullet$ in driving
the Baldwin Effect.  Results shown by \citet{Warner06} provide strong
indications that $M_\bullet$ is in fact a key underlying parameter; 
confirmation with other samples and by other approaches such as PCA would
be desirable.

\section{The Intrinsic Baldwin Effect}

Multi-epoch observations of variable Seyfert galaxies have shown in many
cases an intrinsic Baldwin Effect, i.e. a decrease in emission-line EW as
the source brightens in the continuum.  The data are sufficient to make a
number of generalizing statements about this phenomenon.

A first essential point is that the intrinsic Baldwin Effect
invariably has a steeper slope than the ensemble Baldwin Effect; the
contrast is nicely illustrated in \citeauthor{Kinney90}
(\citeyear{Kinney90}; see also \citeauthor{Pogge92}\citeyear{Pogge92}).  
The two correlations are different and may arise from very different
causes, a distinction that is not always made in the literature (see
OS for further discussion).  A second aspect commonly seen in the
intrinsic effect is {\it curvature} -- the slope displayed in the log
EW vs. log $L$ plane steepens as the source brightens.  An example
appearing recently is the compilation of UV results spanning more than
20 years for NGC~4151 by \citet{Kong06}.  

Ultimately we can understand most aspects of the intrinsic Baldwin
Effect, including its slope and curvature, in terms of photoionization
theory for the BLR, reflected in the responsivity of the line-emitting
clouds as the irradiating continuum varies in intensity.  This issue
has been explored in detail by several authors (e.g.,
\citeauthor{Gilbert03}\citeyear{Gilbert03};
\citeauthor{Goad04}\citeyear{Goad04}; \citeauthor{Korista04}
\citeyear{Korista04}; \citeauthor{Cackett06}\citeyear{Cackett06}).
These studies suggest that in the future the detailed
luminosity-dependent response of the emission lines can serve as a
diagnostic tool for the BLR and its structure.

A specific set of lines with special interest in variable AGNs is the optical
\ion{Fe}{ii} emission.  Two recent studies illustrate the continuing
uncertainty in \ion{Fe}{ii} behavior.  \citet{Wang05} measured optical
\ion{Fe}{ii} emission in the variable NLS1 NGC~4051 and found that the
\ion{Fe}{ii} varied with {\it greater} amplitude than H$\beta$ and the
continuum -- i.e., \ion{Fe}{ii} shows an inverse Baldwin Effect.  In
contrast, \citet{Vestergaard05} reported that the amplitude of
variability for optical \ion{Fe}{ii} in NGC~5548 was only 50\% -- 75\%
that of H$\beta$.  Measuring \ion{Fe}{ii} emission is inherently
difficult due to the large number of features and their substantial
blending.  Further investigations of \ion{Fe}{ii} response would be of
interest given the significant energy carried by these lines and
continuing uncertainty about their excitation.  \citet{Vestergaard05}
have noted that the fact that \ion{Fe}{ii} responds at all to the
continuum already indicates that line fluorescence in a photoionized
plasma apparently dominates over collisional heating in exciting the
observed emission.

\section{The Baldwin Effect in {\em Narrow} Lines}

Information on the luminosity dependence of the narrow emission lines
in AGN ensembles has grown significantly in recent years, and there is
now clear evidence that at least some of the narrow lines display a
Baldwin Effect (e.g., \citeauthor{Boroson92}\citeyear{Boroson92};
\citeauthor{McIntosh99}\citeyear{McIntosh99}; \citeauthor{Croom02}
\citeyear{Croom02}; \citeauthor{Dietrich02}\citeyear{Dietrich02};
\citeauthor{Netzer04}\citeyear{Netzer04}; \citeauthor{Netzer06}
\citeyear{Netzer06}). The best studied feature is [\ion{O}{iii}]
$\lambda$5007, although \citet{Croom02} have compiled extensive results
also for [\ion{Ne}{v}] $\lambda$3426, [\ion{O}{ii}] $\lambda$3727, and
[\ion{Ne}{iii}] $\lambda$3869.  \citeauthor{Croom02} find a Baldwin
Effect for both [\ion{Ne}{v}] and [\ion{O}{ii}], and marginal evidence
for a trend in [\ion{Ne}{iii}].

The narrow emission lines in AGNs originate on vastly larger
dimensions than the BLR size scale, and for this reason if not others
it is quite possible that luminosity dependence in the EWs may have
very different causes for the narrow and broad lines.  One likely
explanation for the Baldwin Effect that is specific to the narrow
lines takes note of the fact that in luminous AGNs, the scale of the
narrow-line region (NLR) may become comparable in size to the entire
galaxy, with the result that the emission measure of the gas cannot
grow further if the luminosity increases; the NLR has run out of gas
\citep[e.g.,][] {Croom02}.  The NLR that is present may also end up
being more highly ionized or partially ejected from the galaxy, which
could result in less emission in the optical forbidden lines.

The reduced narrow-line EWs in luminous AGNs have additional
potentially significant implications.  The existence and numbers of
``Type 2'' QSOs, i.e. lacking broad emission by analogy with Seyfert 2
galaxies, is of interest for understanding the statistics of BLR
obscuration, the true space density of accreting black holes, and the
possibility that some QSOs genuinely lack an intrinsic BLR.  Finding
Type 2 QSOs has proven challenging and often relies on their detection
in the narrow forbidden lines.  Many luminous Type 2 AGNs thus may be
missing from our census if their NLR emission is reduced in relative or
absolute terms, as implied by the Baldwin Effect (\citeauthor{Croom02}
\citeyear{Croom02}; \citeauthor{Netzer06}\citeyear{Netzer06}).  

The NLR Baldwin Effect also has relevance to the possibility of using
nebular lines as tracers of star formation in AGN hosts.  \citet{Ho05}
has noted that the weak [\ion{O}{ii}] emission characteristic of quasars
implies very low levels of star formation if standard conversions between
$L$([\ion{O}{ii}]) and star formation rate (SFR) are applied; if part of the
[\ion{O}{ii}] emission originates in NLR gas powered by the AGN, the
estimated SFR for the remaining fraction is even less.  The fact that
NLR emission appears weak could imply that star formation is suppressed by
the influence of the accretion source, or perhaps that star formation
is confined to dense, shielded regions with low filling factor. In any case,
the Baldwin Effect observed for narrow lines is probably telling us
important things about the AGN host galaxies for luminous systems. 

\section{The X-ray Baldwin Effect}

The dominant line in the hard X-ray spectrum of AGNs is the Fe
K$\alpha$ feature, and there is ongoing debate as to whether this
feature exhibits a Baldwin Effect.  The initial suggestion of such a
trend was made by \citet{Iwasawa93} based on {\it Ginga}
measurements, and \citet{Nandra97} subsequently found confirming results
using {\it ASCA}.  The advent of {\it XMM-Newton} and {\it Chandra}
has substantially increased the quantity and quality of Fe K$\alpha$
measurements for AGNs, and several papers using data from these
telescopes have appeared, supporting the existence of a Baldwin Effect
in most cases (\citeauthor{Page04} \citeyear{Page04},
\citeyear{Page05}; \citeauthor{Zhou05}\citeyear{Zhou05}; 
\citeauthor{Jiang06}\citeyear{Jiang06}) but not all
\citep{Jimenez05}.  At this conference Stefano Bianchi presented
measurements for a new, expanded sample that provide support for the
existence of a Baldwin Effect for Fe K$\alpha$.

Several explanations have been put forward for the X-ray Baldwin Effect.
Since the Fe K$\alpha$ line originates via reprocessing involving
photoelectric absorption in high column density gas, one possibility
is that the covering factor of such gas is reduced in higher
luminosity sources.  A clue to the nature of the reprocessor is
provided by the line profile.  The {\it XMM} results indicate that the
emission is mostly narrow (i.e., nonrelativistic, with widths of 
$10^3 - 10^4$ km s$^{-1}$), suggesting that the reprocessor may be the
BLR, the outer accretion disk, or the circumnuclear torus.  If the
emission originates from the disk, a theoretical prediction consistent
with a Baldwin Effect is that luminous sources driven by large
accretion rates will produce less Fe K$\alpha$ emission, as the disk
becomes sufficiently ionized that much of the Fe is fully stripped of
electrons (e.g., \citeauthor{Matt93} \citeyear{Matt93};
\citeauthor{Nandra97}\citeyear{Nandra97}).  If the line originates
from the torus, the observed trend would be a natural consequence of
the ``receding torus'' model in which the inner boundary of the torus
moves outward as luminosity increases, so that the covering factor
is reduced if the torus scale height remains constant \citep{Lawrence91}.
If Fe K$\alpha$ is produced substantially in the BLR, the same medium
produces the UV/optical lines that show the original Baldwin Effect;
a luminosity-dependent covering factor may account for part of the
trend, but the inverse correlation seen in some lines (notably H$\beta$,
\S 2.1) makes this explanation less satisfying.

Other explanations that do not involve a luminosity dependence in the
reprocessor structure are possible, however.  \cite{Jiang06} have
argued that the X-ray Baldwin Effect is substantially driven by
radio-loud sources.  In these objects we may be observing an extra
X-ray continuum component from a relativistic jet that does not
participate in Fe K$\alpha$ production, so that when the jet component
is strong, the source appears more luminous while EW(K$\alpha$)
appears weak \citep[see also][]{Jimenez05}.  \cite{Jiang06} find that
radio-quiet AGNs considered alone display a weak Baldwin Effect in
Fe K$\alpha$ that can be understood as a consequence of variability;
if the light-crossing time for the region emitting Fe K$\alpha$ exceeds
the variability timescale for the X-ray continuum, a Baldwin Effect
naturally occurs as the continuum goes up and down, and individual sources
in an ensemble are measured at random phases.  \citet{Miniutti04}
note that light-bending effects near a rotating black hole may act
to accentuate the observed trend.

The observational situation remains dynamic, as illustrated by S. Bianchi's
results presented at this meeting showing a significant X-ray Baldwin Effect
for a large sample restricted to radio-quiet objects.  Discussion
of the X-ray Baldwin Effect and issues related to Fe K$\alpha$ emission
were presented by Giorgio Matt and Tahir Yaqoob, and the reader is referred
to their contributions for further details.

\section{Conclusions}

Considerable progress has been made in recent years in our
understanding of the Baldwin Effect and related correlations.
Luminosity-dependent EWs are clearly seen in both the broad and narrow
lines for AGN ensembles.  There are strong indications that the broad-line
trends may actually derive from more fundamental correlations with
other parameters such as $M_\bullet$ or $L/L_{Edd}$, although a
consensus in the community on this point has not yet emerged.
Integrating studies of the Baldwin Effect into more comprehensive PCA
analyses is a necessary step to advance this field of inquiry.
NLS1 galaxies appear to be outliers from many of the correlations and
there is a need to address why this is the case.  A theme that has
surfaced repeatedly through the Baldwin Effect's history is that the
details of sample selection remain very important.  Further progress
is possible and clearly desirable.

\end{document}